\documentstyle[aps,prb]{revtex}
\newcommand{\ds}{\displaystyle}
\newcommand{\noi}{\noindent}
\newcommand{\non}{\nonumber}
\newcommand{\ii}{{\rm i}}
\newcommand{\e}{\varepsilon}
\newcommand{\k}{{\mbox{\boldmath$k$}}}
\newcommand{\x}{{\mbox{\boldmath$x$}}}
\newcommand{\r}{{\mbox{\boldmath$r$}}}
\newcommand{\y}{{\mbox{\boldmath$y$}}}
\newcommand{\q}{\mbox{\boldmath$q$}}
\newcommand{\LQ}{\mbox{\boldmath$Q$}}
\newcommand{\SLQ}{\mbox{\boldmath$Q$}}
\newcommand{\sk}{{\mbox{\boldmath$k$}}}
\newcommand{\sq}{\mbox{\boldmath$q$}}
\newcommand{\sx}{{\mbox{\boldmath$x$}}}
\newcommand{\be}{\begin{eqnarray}}
\newcommand{\en}{\end{eqnarray}}
\newcommand{\dd}{{\rm d}}
\newcommand{\JPSJ}{J. Phys. Soc. Jpn.}

\newcommand{\PRL}{ Phys. Rev. Lett.}
\newcommand{\PRB}{ Phys. Rev. {\bf  B}}

\newcommand{\PROG}{Prog. Theoret. Phys}
\begin{document}
\title{Quantum phase transitions and collapse of the Mott gap 
in the $d=1+\epsilon$ dimensional half-filled Hubbard model}
\author{Jun-ichiro Kishine\thanks{E-mail:kishine@ims.ac.jp}}
\address
{Department of Theoretical Studies,
Institute for Molecular Science,
Okazaki 444-8585, Japan}
\date{\today}
\maketitle
\begin{abstract}
We  study the low-energy asymptotics of the half-filled Hubbard model
with a  circular Fermi surface
in $d=1+\epsilon$ continuous dimensions,
based on the one-loop renormalization-group (RG) method.
Peculiarity of  the $d=1+\epsilon$ dimensions
is  incorporated  through  the  mathematical structure of the elementary 
particle-partcile (PP) and particle-hole (PH) loops:
infrared logarithmic singularity of the PH loop is smeared for $\epsilon>0$.
The  RG flows indicate that a quantum phase transition (QPT) 
 from a metallic phase to the Mott insulator  phase
occurs at a finite  on-site Coulomb repulsion $U$ for  $\epsilon>0$.
We also discuss effects of randomness.
\end{abstract}

\section{INTRODUCTION}
Correlation-driven metal-insulator transition (MIT)  
in the   Hubbard model has been a basic problem in condensed matter physics.\cite{Imada98}Central to this issue is the problem of quantum fluctuation controlled
by the on-site Coulomb repulsion $U$, the band width $W$, and the carrier 
concentration $n$.
In the half-filled ($n=1$) case,
 the zero temperature band width control MIT occurs at 
a  critical ratio of $\bar U=U/W$, ${\bar U}_{\rm c}$.
The exact solution of the Hubbard model is available only in
 $d=1$ dimension, where
the half-filled ground state is always an insulator with a finite charge excitation gap (Mott gap)
at $\bar U > \bar U_{\rm c}=0$.\cite{LiebWu68}  
In $d=2$, it is believed that there exists a finite ${\bar U}_{\rm c}$
 except in the case of the perfect nesting where ${\bar U}_{\rm c}=0$.\cite{Hirsh85} 
 In the case of the $d=2$ half-filled Hubbard model with the nearest neighbor 
  and the second nearest
 neighbor hoppings, 
a Monte Carlo simulation,\cite{LinHirsh87} a Hartree-Fock approximation, and
 the Gutzwiller approximation\cite{KondoMoriya96} all indicate
 that $\bar U_{\rm c}$ is finite.
In $d=\infty$, the dynamical mean field approach\cite{Metzner89}
 implies that the quasiparticle spectral weight
in the vicinity of the Fermi surface vanishes continuously as $U$ approaches  
a critical value $\bar U_{\rm c}\sim 1$  from below.\cite{Georges96}
The filling control MIT,  which occurs as the carrier concentration
approaches half filling, has also been extensively studied.\cite{Imada98}

In the $d=1$ Hubbard model at half filling, low-energy asymptotics  
is also well understood in terms of the renormalization-group (RG) flow 
of two-particle scattering strengths, $g_1$, $g_2$, and $g_3$,
which correspond to the backward, forward and $2k_F$-umklapp scatterings, respectively.\cite{Kimura75,Solyom}
The RG scheme in $d=1$ is based on infrared logarithmic singularities of
 elementary particle-particle (PP) and  particle-hole (PH) loops which have
  the same magnitude and opposite signs.
In this context, the source of the Mott gap  is the umklapp scattering which becomes a relevant perturbation for $U>0$.
The RG flow also indicates that the charge stiffness reaches zero during the renormalization process and 
 the system becomes an insulator. Thus, the RG-based scenario is
in perfect agreement with the exact solution.

Dimensionality effects on the $d=1$ Mott insulator phase were  phenomenologically
treated
 by cutting off either of the PP or the PH loop below some characteristic temperature.\cite{Emery82,Yonemitsu97} 
In the case of  weakly-coupled chains,
dimensional crossovers caused by an interchain hopping $t_\perp$ 
 have   been studied by treating $t_\perp$ perturbatively 
 and  assuming that
the scaling procedure in the one-dimensional regime at 
high energy  scales ( $\omega\gg t_\perp$)  remains valid 
down to the crossover energy scales.\cite{Bourbonnais93,Suzumura98,Kishine99}
Recently,  the filling control MIT in 
 coupled Hubbard chains with  infinitely large
coordination numbers was also studied.\cite{Fujimoto99}
However, these attempts have not clarified the dimensionality effects
on the Mott gap, because 
feedback effects of the interchain processes on the  Mott gap have been missing.

The RG method is  straightforwardly extended
to  the case of $d=1+\epsilon$ $(0<\epsilon\ll 1)$  dimensions.\cite{Ueda84}  
In this case, only the PP loop    remains  logarithmically singular,
 while  the PH loop is smeared for $\epsilon>0$.\cite{Metzner97}
By taking this fact into account
from the beginning of renormalization processes, 
the feedback effects on the   Mott gap may be incorporated.
In this paper, by using   the one-loop RG method,
we examine dimensionality effects  on the RG flow of the umklapp process and discuss
possible QPTs in the
 Hubbard model with a circular Fermi surface in $d=1+\epsilon$ dimensions.

If randomness exists,  a QPT between
the Mott insulator and a randomness-driven Anderson insulator might arise.
In the case of $d=1$ at half filling, based on the RG method, 
Fujimoto and Kawakami\cite{Fujimoto96} found 
 that sufficiently strong random forward scattering 
destroys the Mott gap.
 Recently, Ohtsuka {\it et al.}\cite{OtsukaMoritaHatugai98,OhtsukaHatsugai99} studied  
 the half-filled Hubbard model containing site randomness
 by using the quantum Monte Carlo technique and found that
  the strong randomness destroys the Mott gap.
In this paper, we  also discuss  the QPTs in
the half-filled random Hubbard model  in $d=1+\epsilon$ dimensions.
 
This paper is organized as follows:
In Sec.~I$\!$I we introduce the $g$-ology effective action,
derive the one-loop RG equations  and
discuss possible QPTs in the absence of randomness.
The effects of randomness 
are discussed in Sec.~I$\!$I$\!$I, followed by concluding remarks
in Sec.~I$\!$V.

\section{HALF-FILLED HUBBARD MODEL IN ${\mbox{$d$}}=1+\epsilon$ DIMENSIONS}
In this section, 
we study 
interplay of electron correlation and dimensionality effects
in the half-filled
Hubbard model with a circular Fermi surface in
$d=1+\epsilon$ dimensions.

\subsection{Effective action}

We start with the  effective  action,  
\begin{eqnarray}
S_{\rm Hubbard}&=&\sum_{\sigma}\int_{-\infty}^{\infty}{d\e\over 2\pi}
\ds\int{d^d\k\over (2\pi)^d}
{\cal G}^{-1}(\k,\ii\e)  c_{\sigma}^{\ast}(K) c_{\sigma}(K)\non\\
&-&
{\pi v_{F} \over 2}\sum_{\sigma,\sigma'}
\prod_{i=1}^{4}
\int_{-\infty}^{\infty}{d\e_i\over 2\pi}
\int{d^d\k_{i}\over (2\pi)^d}
\delta(\e_4\!+\!\e_3\!-\!\e_2\!-\!\e_1)
\delta(\k_4\!+\!\k_3\!-\!\k_2\!-\!\k_1\!
-\!{\mbox{\boldmath$G$}})\non\\
&& \,\,\,\,\,\,\,\,\,\,\,\,\,\,\,\,\,\,\,\,\,\,\,\,\,\,\,\,\,\,\,
\ds
g^{\sigma\sigma'}_{\sk_1,\sk_2,\sk_3}
c_{\sigma}^{\ast}(K_4)  
c_{\sigma'}^{\ast}(K_3)  
c_{\sigma'}(K_2)  
c_{\sigma}(K_1),\label{eqn:Hub}
\end{eqnarray}
where $c^\ast_{\sigma}(K)$ and $c_{\sigma}(K)$ are  the Grassmann variables 
representing  the   electron with the spin $\sigma$  and
$K=(\k,\e)$ with  $\k$ and $\e$ being a $d$-dimensional momentum and a Fermion thermal frequency, respectively.
The non-interacting one-particle propagator is given by
\be
{\cal G}^{-1}(\k,\ii\e)=\ii\e-\xi(\k).
\en
The one-particle  dispersion is $\xi(\k)=(\k^2-k_F^2)/2m$ and
the Fermi surface is a $d$-dimensional sphere $\mid\k\mid=k_F$. 
Since we consider only  the energy scale which is much smaller than the Fermi energy,
 we linearize  one-particle dispersion as
\be
\xi(\k)=v_F(\mid\!\k\!\mid-k_F), \label{eqn:1Pdispersion}
\en
where 
$v_F$ is the Fermi velocity.
The band width cut off $E_0$ is introduced and the one-particle processes are restricted
to $-E_0\leq\xi(\k)\leq E_0$.
In the two-particle scattering part of $S_{\rm Hubbard}$,
$\mid\!{\mbox{\boldmath$G$}}\!\mid=0$ and $\mid\!{\mbox{\boldmath$G$}}\!\mid=4k_F$ for
the normal and the umklapp processes, respectively.
It is reasonable to assume that the two-particle scattering processes which enter  the  
 RG equations in $d=1+\epsilon$ are,
as in the case of $d=1$,
the backward scattering with large momentum transfer $\mid\!\k_3-\k_2\!\mid\sim 2k_F$,
the forward scattering with small momentum transfer $\mid\!\k_3-\k_2\!\mid\sim 0$,
and the $2k_F$-umklapp  scattering characterized by $\mid\!\k_4\!+\!\k_3\!-\!\k_2\!-\!\k_1\!\mid\sim 4k_F$. 
We represent the corresponding scattering
vertices  in Figs.~1(a)-(c).
Dimensionless scattering strengths for these processes are denoted by
$g_{1}$, $g_{2}$, and $g_{3}$, respectively.
Unrenormalized scattering strengths are related to the on-site Coulomb repulsion, $U$, as\cite{Solyom}
\be
g_{1;0}=g_{2;0}=g_{3;0}=U/\pi v_F\equiv\tilde U.
\en

\subsection{The elementary particle-particle and particle-hole loops}

The   peculiarity of  $d=1+\epsilon$ dimensions is incorporated only through
the integration measure, $\ds\int{d^d\k/(2\pi)^d}$.
For our purpose here, it is sufficient to integrate over $k=\mid\!\k\!\mid$
and the angle $\theta$ spanned by
$\k$ and another fixed momentum. Then we can use\cite{Wilson72}
\be
\ds\int{d^d\k\over (2\pi)^d} (\cdots)
={S_{d-1}\over(2\pi)^d}\int k^{d-1}dk\int_{0}^\pi d\theta (\sin\theta)^{d-2}(\cdots),
\en
where  
$S_d=2\pi^{d/2}/\Gamma(d/2)$ is the surface area of the $d$-dimensional
unit sphere.

As is well known,
in any dimension, the real part of the elementary PP loop [Fig.~1(d)]
at the zero total momentum, 
\be
\Delta_{0}(\omega)=\ds\int{d^d\k\over (2\pi)^d}
{\Theta(\xi_{-\sk})-\Theta(-\xi_{\sk})\over \omega -\xi_{-\sk}-\xi_{\sk}+\ii 0^{+} },
\label{eqn:real}
\en
with $\Theta(x)$ being the step function, exhibits an infrared logarithmic singularity of the form,\cite{Ueda84,Metzner97}
\be
\Re\Delta_{0}(\omega) =-{S_d\over (2\pi)^d}{k_F^{d-1}\over v_F}\log(\omega/E_0).
\en
In $d=1+\epsilon$, we  obtain
\be
\Re\Delta_{0}(\omega) \sim -{1\over 2\pi v_F} \log(\omega/E_0),
\en
which exactly reproduces the result in $d=1$.

On the other hand,  the real part of the elementary PH loop at $2k_F$ momentum transfer[Fig.~1(e)], 
\be
\Pi_{2k_F}(\omega)=\ds\int{d^d\k\over (2\pi)^d}
{\Theta(\xi_{\sk+\SLQ})-\Theta(\xi_{\sk})\over \omega -\xi_{\sk+\SLQ}+\xi_{\sk}+\ii 0^+},
\label{eqn:ret}
\en
with $\mid\!\! \LQ\!\!\mid=2k_F$, unlike the case of $d=1$,
no longer exhibits an infrared singularity for $d>1$ and 
in $d=1+\epsilon$, takes the form,
\be
\Re\Pi_{2k_F}(\omega)\sim{1\over 2\pi v_F}
\left[{ \tilde\omega^{\epsilon/2}\over \epsilon/2}+C_\epsilon\right],\label{eqn:phloop}
\en
where $\tilde\omega=\omega/2 v_F k_F$ and $C_\epsilon$ is a constant independent 
of $\omega$. 
Although this form has already been suggested in Ref.~[16],
 we confirm, in the appendix,  
 that the $\omega$-dependent term  in (\ref{eqn:phloop}) 
is uniquely determined. 

\subsection{One-loop renormalization}

One-loop renormalization of the scattering strengths comes from
the vertex correction diagrams represented in Figs.~2(a) and 2(b) 
for the normal [$g_1$, $g_2$] and
umklapp [$g_3$] processes, respectively.
The renormalized scattering strength, $g'_1$, $g'_2$, and $g'_3$, are thus given by
\be
g'_1&=&g_1+{g_1g_2}\ln{\omega\over E_0}-(g_2-g_1)g_1\pi(\omega)\label{vc1},\\
g'_2&=&g_2+{1\over2}({g_1^2+g_2^2})\ln{\omega\over E_0}-{1\over2}({g_2^2+g_3^2})\pi(\omega),\\
g'_3&=&g_3+{g_3}(g_1-2g_2)\pi(\omega)\label{vc3}, 
\en
where
$
\pi(\omega)=2\pi v_F\Re\Pi_{2k_F}(\omega).
$
Fig.~2(a-1) contains the PP loop and all other diagrams contain the PH loop.
In particular, renormalization of the umklapp process comes from the PH loop only.
By differentiating equations (\ref{vc1})-(\ref{vc3}) with respect to the scaling parameter\be
l=\ln(E_0/\omega),
\en 
we obtain the RG equations 
\begin{eqnarray}
{\dd  g_{1}\over  \dd l} &=&
-g_{1}g_{2} + (g_{2}-g_{1} )  g_{1}\lambda_l,\label{eqn:rgeqg1}\\
{\dd  g_{2} \over \dd l} &=&-(g_{1}^2+g_{2}^2)/2 +(g_{2}^2+g_{3}^2)\lambda_l/2  ,\\
{\dd  g_{3} \over \dd l} &=&-g_{3} (g_{1}-2g_{2} ) \lambda_l  \label{eqn:rgeqg3}.
\end{eqnarray}
The PH loop gives rise to
the smooth cutoff,\cite{Ueda84} 
\be
\lambda_l\equiv  \left|{\partial \over \partial l}\pi(\omega)\right|
=\left({E_0\over 2k_Fv_F}\right)^{\epsilon/2}\exp[-\epsilon l/2]
\sim \exp[-\epsilon l/2],
\en
where the ratio of the two cut off energy scales, $E_0/2k_Fv_F$,
 is of the  order of unity.
In the absence of the umklapp process, 
the RG equations obtained here reproduces those in Ref.~[16]. 

\subsection{Renormalization-group flow and QPT}
In the $d=1$ half-filled Hubbard model, 
the  charge degrees of freedom  are governed by the combination of $(g_{3},G=g_1-2g_2)$  
 with the flow lines $(G-{\rm const.})^2-{g_{3}}^2={\rm const}$. 
For {\it any finite} $U>0$, they are scaled to 
$(g_3^\ast=\infty,G^\ast=-\infty)$, which
 implies the  Mott gap opens due to the relevant umklapp scattering. 
 This RG flow also indicates that the charge stiffness
 $K_\rho=\sqrt{(1+G)/(1-G)}$ reaches zero during the renormalization process and 
 the system becomes an insulator. 
 To complement this scenario, it is useful to map
the charge sector of the $d=1$ half-filled
Hubbard model  onto the (1+1)-dimensional
 sine-Gordon model by using the bosonization technique.
 The sine-Gordon action is 
of the form:\cite{Solyom}
\begin{eqnarray}
S_{\rm SG}=\int d^2 \r \left\{
{u_\rho\over 2}(\nabla_\r \Psi_\rho(\r))^2
-{2g_3\over (\pi\alpha)^2} \cos [\sqrt{8\pi K_\rho}\Psi_\rho(\r)]\right\}
\label{eqn:SG},
\end{eqnarray}
where $\r=(x,\tau)$ represents (1+1)-dimensional space-time coordinates,
$\Psi_\rho(\r)$ represents a charge boson (holon) field,
$u_\rho$ is the holon velocity, and $\alpha$ is a short distance cut off.
By applying the RG method directory to
the sine-Gordon model,\cite{Weigmann78} we  obtain the  RG equations
\begin{eqnarray}
{\dd g_3\over \dd l}&=&2(1-K_\rho) g_3,\label{bosong3}\\
{\dd K_\rho \over \dd l}&=&-{8a\pi^2\over \Xi_l^4} g_3^2 K_\rho^3,\label{bosonK}
\end{eqnarray}
where $a=\int_0^1d\rho\rho^3J_0(\rho)$ and
$\Xi_l=\Xi_0e^{-l}$ is the space-time cutoff.
We thus see  that
$g_3$ becomes relevant for the initial condition $K_{\rho;0}<1$
[corresponding to $U>0$ for the Hubbard model]
and then, accordingly, the charge stiffness $K_\rho$ is scaled to zero.

Now we consider the case of $d=1+\epsilon$ dimensions.
In Figs.~3(a) and (b) are shown the RG flows for $\tilde U=U/\pi v_F=0.08$ and $0.04$,
respectively, in $d=1.1$.
In Fig.~3(c), we show the RG flows in terms of $g_3$ and $G=g_1-2g_2$ for various
$\tilde U$ in $d=1.1$. It is found that there exists a critical value of
$\tilde U$, $\tilde U_{\rm c}=0.0588$.
For $\tilde U>\tilde U_{\rm c}$, the RG flows exhibit runaway trajectories
toward $(g_3^\ast=\infty, G^\ast=-\infty)$ [shaded region in Fig.~3(c)], which
implies the  Mott gap opens  at the low-energy limit
just as in the case of $d=1$.
The initial values of $g_3$ and $G$ at $l=0$ correspond to the points on 
the line, $G=-g_3$, represented by a broken line in Fig.~3(c).
In the $d=1$ half-filled Hubbard model, 
$g_3$ and $G$ flow along this line (denoted by \lq\lq 1D Hubbard\rq\rq) 
for any $U>0$.

On the other hand, the RG flows approach the  fixed points,
$(g_3^\ast={\rm const.}, G^\ast=0)$
for $\tilde U<\tilde U_{\rm c}$. 
Marginal behavior of $g_3$ is in accordance with $G=g_{1}-2g_{2}\to 0$ [see equation (\ref{eqn:rgeqg3})] as $l\to \infty$.
The  smooth cutoff,
$\lambda_l=\exp[-\epsilon l/2]$, in eq. (\ref{eqn:rgeqg3}) 
causes suppression of $g_3$  during the renormalization process. This suppression becomes more  conspicuous for larger $\epsilon$. 
The fixed point $G^\ast=0$, corresponding to 
the non-interacting value of the charge stiffness, $K_\rho^\ast=1$, 
implies  that the Mott gap collapses and the system becomes metallic at the low-energy limit. Thus  a QPT from the metallic phase to the Mott insulator  phase
may occur at
$\tilde U=\tilde U_{\rm c}$.

Within the RG-based scheme,
it remains debatable how the marginal behavior of $g_3$ for $\tilde U>\tilde U_{\rm c}$
modifies the ground state property. Regarding this point, recently the density matrix
renormalization group (DMRG) method was applied to  three Hubbard chains coupled via
the interchain one-particle hopping $t_\perp$.\cite{Yonemitsu99}   As a result, it was found that
the Mott gap decreases as $t_\perp$ increases. 
This numerical result  strongly supprots  the RG-based view given here.   

We here 
give qualitative discussion on the magnitude of the Mott gap.
There is no tractable
method to quantitatively obtain the  magnitude of the Mott gap in $d>1$.
However, the magnitude of the Mott gap is qualitatively
given by the energy scale, $\omega_{\rm gap}=E_0e^{-l_{\rm gap}}$, at which the umklapp scattering
strength exceeds unity, $g_3=1$ [see Fig.~3(a)].\cite{Suzumura98,Kishine99}
To see how $\omega_{\rm gap}$ reproduces  the Mott gap, we compare
 the $\tilde U$-dependence of the exact Mott gap in $d=1$,\cite{Takahashi99}
\be
\Delta_{\rm exact}(\tilde U)={2v_F\over\pi^2{\tilde U}^2}\int_{1}^\infty d\eta{\sqrt{\eta^2-1}\over\sinh[\eta/\tilde U]},
\en
with that of $\omega_{\rm gap}$ in $d=1$.
There is arbitrariness in specification of the linearized bandwidth $E_0$.
In Fig.~4, we show the case for $E_0=0.4v_F$, where $\omega_{\rm gap}$ reproduces 
 $\Delta_{\rm exact}(\tilde U)$ well 
at least for a weak $\tilde U$ where the weak coupling RG scheme  is valid.

In Fig.~5, we show a low-energy asymptotic phase diagram.
We also show $\omega_{\rm gap}/E_0$ as a function of $d$ and $\tilde U$.
For a fixed dimension, the low-energy asymptotic phase corresponds to
a metal and the Mott insulator phases  for
 $\tilde U<\tilde U_{\rm c}$ and  $\tilde U>\tilde U_{\rm c}$, respectively.
The critical value, $\tilde U_{\rm c}$, increases as $d$ increases, suggesting
a larger $U$ is required for the Mott gap to open  as  the dimension increases. 
Accordingly, for a fixed $\tilde U$,
$\omega_{\rm gap}/E_0$ decreases with increasing dimensions
 and disappears at some
critical dimension, $d_{c}$ [ for example, $d_{c}=1.345$
for $\tilde U=0.2$ ].

Apparently $\omega_{\rm gap}/E_0$ grows continuously at $\tilde U=\tilde U_{\rm c}$ for $d\stackrel{<}{\sim}1.4$.
However, 
the transition is always discontinuous by the following reason. 
As is seen from Fig.~3(c),  $g_3$ exceeds unity during the renormalization process
and gives a  finite value of $\omega_{\rm gap}/E_0$ even 
for $\tilde U<\tilde U_{\rm c}$.
Nevertheless,
$\omega_{\rm gap}/E_0$ has no physical interpretation for 
$\tilde U<\tilde U_{\rm c}$, since the charge stiffness 
is scaled to the non-interacting value.
Therefore,  $\omega_{\rm gap}/E_0$ grows discontinuously
at $\tilde U=\tilde U_{\rm c}$. 
The magnitude of discontinuity at the transition
point becomes more conspicuous as $d$ increases.
At present, it is not clear whether 
the discontinuity is an artifact of the one-loop RG method.
To elaborate  on this point would require
 the {\it two-loop} RG analysis, which is too involved a subject to be 
 treated here.

We here comment on the relevance of the present study to the $d=2$  lattice 
Hubbard model.
In this paper, we have studied only the case with a {\it circular} Fermi surface. 
Accordingly, our results
are not smoothly connected to the $d=2$ square lattice Hubbard model.
In particular, in the case of the perfect nesting, 
the van-Hove singularity at $(\pm \pi,\pm\pi)$ points gives rise to
the \lq\lq log-square\rq\rq singularity  of the elementary
PH loop at the momentum transfer $\LQ=(\pi,\pi)$.
Then  the Hartree-Fock solution gives the gap\cite{Hirsh85}
$
\Delta\sim te^{-2\pi\sqrt{t/U}},
$
which indicates
that  the ground state of the system is always an insulator for a finite $U$.
This conclusion has  been strongly supported by numerical studies.\cite{Imada98,Hirsh85}
However possibility of the MIT  at a finite $U$ in the half-filled lattice Hubbard model 
with a Fermi surface of various geometries
 has been  open to question.
 In the case of the $d=2$ half-filled Hubbard model with the nearest neighbor 
  and the second nearest
 neighbor hopping integrals ( $t$ and $t'$, respectively ),
a Monte Carlo simulation,\cite{LinHirsh87} a Hartree-Fock approximation and
 the Gutzwiller approximation including the antiferromagnetism\cite{KondoMoriya96}
 give finite critical values, $U/t=2.5$, 2.064, and 3.902, respectively, for
 $t'/t=0.2$. These results are consistent with the present findings that,
{\it in the case of a circular Fermi surface, 
 the MIT occurs at a finite $U$ for $\epsilon>0$.} 
 
\section{EFFECTS OF RANDOMNESS}
In this section, 
we study 
interplay of electron correlation, randomness and dimensionality effects
in the $d=1+\epsilon$ dimensional random Hubbard model at half filling.\cite{progress}

\subsection{Effective action for quenched randomness}

The  action for the scattering processes by the random potentials, 
\begin{eqnarray}
S_{\rm random}&=&
-\sum_{\sigma}\int d^d\x \int
 d\tau v(\x)c^\ast_\sigma(\x,\tau) c_\sigma(\x,\tau)\non\\
&=&
-\sum_{\sigma}\int {d^d\k\over (2\pi)^d} \int {d^d\q\over (2\pi)^d}
\int d\tau v(\q)c^\ast_\sigma(\k+\q,\tau) c_\sigma(\k,\tau),
\end{eqnarray}
is added to the Hubbard action (\ref{eqn:Hub}).
Here, $c_\sigma(\x,\tau)=\sum_{\sk}e^{\ii\sk\cdot\sx}c_\sigma(\k,\tau)$ with $\tau$ being an imaginary time and
$v(\x)=\sum_{\sq}e^{\ii\sq\cdot\sx}v(\q)$ is  a random   potential at the position  $\x$.
We assume that the random  scattering processes which enter  the  
 RG equations in $d=1+\epsilon$ are,
as in the case of $d=1$,\cite{Giamarchi88}
characterized by
real and complex random fields $\eta(\x)$ and $\xi(\x)$ 
for the  forward scattering with small momentum transfer $\mid\!\q\!\mid\sim 0$ 
and the backward scattering with large momentum transfer $\mid\!\q\!\mid\sim 2k_F$,
respectively, due to the random potential.
The random potentials are assumed to be governed by  Gaussian distributions,
\begin{eqnarray}
P_\eta &\propto&\exp\left[-D_\eta\int d^d\x\eta(\x)^2\right], \\
P_\xi&\propto&\exp\left[-D_\xi\int d^d\x\xi(\x)\xi^\ast(x)\right],
\end{eqnarray}
which lead to
\begin{eqnarray}
\langle\eta(\x)\eta(\y)\rangle_{\rm random} &=& {D_\eta\over 2 }\delta(\x-\y),\\
\langle\xi(\x)\xi^\ast(\y)\rangle_{\rm random} &=& {D_\xi\over 2 }\delta(\x-\y), 
\end{eqnarray}
where $D_\eta=(\pi N_F \tau_\eta)^{-1}$ and $D_\xi=(\pi N_F \tau_\xi)^{-1}$ with 
$\tau_{\eta,\xi}$  and $N_F$ being the elastic-scattering mean free time and  
 the  one-particle density of states, respectively.

We consider the quenched randomness  where  averaging the free energy 
 is accomplished by means of the replica trick which is based on the identity
\begin{eqnarray}
\ln Z=\lim_{N\to 0}{Z^N-1\over N}.
\end{eqnarray} 
We introduce $N$ identical replicas of the system labeled by the index $\alpha$.
Then, by using the path-integral
 representation of the partition function, we have 
\begin{eqnarray}
Z^N=\int \prod_{\alpha=1}^N {\cal D}c^{\alpha\ast}{\cal D}c^\alpha \exp\left[\sum_{\alpha=1}^N{S^\alpha}\right],
\end{eqnarray}
where $S^\alpha=S_{\rm Hubbard}^\alpha+S_{\rm random}^\alpha$
 is the total action  and $\cal D$ symbolizes the measure 
  over the fermionic Grassmann variables $c^{\alpha\ast}$ and $c^\alpha$ 
  depending on a replica index $\alpha$.
 The replica trick consists of performing the Gaussian ensemble average $\langle Z^N\rangle_{\rm random}$  for integer $N$,
continuing the result analytically to real $N$, and taking the limit $N \to 0$. 
We thus obtain  
\begin{eqnarray}
\langle Z^N\rangle_{\rm random} =
\int d\eta P_\eta \int d\xi d\xi^\ast P_\xi
\int \prod_{\alpha=1}^N {\cal D}c^{\alpha\ast}{\cal D}c^\alpha  \exp\left[\sum_{\alpha=1}^N{S^\alpha}\right] 
\equiv\int \prod_{\alpha=1}^N {\cal D}c^{\alpha\ast}{\cal D}c^{\alpha} \exp[\sum_{\alpha=1}^N\tilde S^\alpha ],
\end{eqnarray}
where the random scattering parts contained in $\tilde S^\alpha$
are given by:\cite{BelitzKirkpatrick94}
\begin{eqnarray*}
&&{D_\eta\over 4}
\sum_{\beta=1}^N\sum_{\sigma,\sigma'}
\int d\tau_1\int d\tau_2
\int {d^d\k\over (2\pi)^d}\int {d^d\k'\over (2\pi)^d}  
c_{\sigma}^{\alpha\ast}(\k,\tau_1) c^{\alpha}_{\sigma}(\k,\tau_1)  
c_{\sigma'}^{\beta\ast}(\k',\tau_2) c^{\beta}_{\sigma'}(\k',\tau_2) \non
\\
&+&{D_\xi\over 4}
\sum_{\beta=1}^N\sum_{\sigma,\sigma'}
\int d\tau_1\int d\tau_2
\int {d^d\k\over (2\pi)^d}\int {d^d\k'\over (2\pi)^d}  
c_{\sigma}^{\alpha\ast}(\k+\LQ,\tau_1) c^{\alpha}_{\sigma}(\k,\tau_1)  
c_{\sigma'}^{\beta\ast}(\k'-\LQ,\tau_2) c^{\beta}_{\sigma'}(\k',\tau_2),
\end{eqnarray*}
where  $\mid\!\LQ\!\mid\sim 2k_F$.

We here change imaginary time variables $\tau_1$ and $\tau_2$ into
$\Delta\tau=\tau_1-\tau_2$ and $ \tau=(\tau_1+\tau_2)/2$. In 
the integration  over $\Delta\tau$, we introduce a short distance cutoff 
$\Lambda$
and
 keep only the region, $v_F \mid \Delta \tau\mid \leq \Lambda$,  which  couples
the two-particle scattering processes
and contribute to the RG equations. 
Then, 
the
random forward and backward scattering  parts are written as
\begin{eqnarray}
\tilde S^\alpha_{\rm random}&\sim&{D_{\eta} \Lambda\over 2v_F} 
\sum_{\sigma,\sigma'}
\sum_{\beta=1}^N
\int {d^d\k\over (2\pi)^d}\int {d^d\k'\over (2\pi)^d}  
\prod_{i=1}^4
\int_{-\infty}^{\infty}{d\e_i\over 2\pi}
\delta(\e_4\!+\!\e_3\!-\!\e_2\!-\!\e_1)\non\\
&&\,\,\,\,\,\,\,\,\,\,\,\,\,\,\,\,\,\,\,\,\,\,\,\,\,\,\,\,\,\,\,\,\,\,\,\,\,\,\,\,\,\,\,\,\,\,\,\,\,\,\,\,\,\,
c_{\sigma}^{\alpha\ast}(\k,\e_4) 
c_{\sigma'}^{\beta\ast}(\k',\e_3) 
c^{\beta}_{\sigma'}(\k',\e_2)
c^{\alpha}_{\sigma}(\k,\e_1)  
\\
&-&{D_{\xi} \Lambda\over 2v_F} 
\sum_{\sigma,\sigma'}
\sum_{\beta=1}^N
\int {d^d\k\over (2\pi)^d}\int {d^d\k'\over (2\pi)^d}  
\prod_{i=1}^4
\int_{-\infty}^{\infty}{d\e_i\over 2\pi}
\delta(\e_4\!+\!\e_3\!-\!\e_2\!-\!\e_1)\non\\
&&\,\,\,\,\,\,\,\,\,\,\,\,\,\,\,\,\,\,\,\,\,\,\,\,\,\,\,\,\,\,\,\,\,\,\,\,\,\,\,\,\,\,\,\,\,\,\,\,\,\,\,\,\,\,
c_{\sigma}^{\alpha\ast}(\k+\LQ,\e_4) 
c_{\sigma'}^{\beta\ast}(\k'-\LQ,\e_3) 
c^{\alpha}_{\sigma}(\k,\e_2)  
c^{\beta}_{\sigma'}(\k',\e_1),
\label{eqn:randombw}
\end{eqnarray}
where
$
c^\alpha_\sigma(\x,\tau)=T^{1/2}\sum_{\k}e^{\ii(kx-\varepsilon \tau)}c_\sigma^\alpha({\k},\e).
$
The actions for the
random scatterings   
inside the same replica [$\beta=\alpha$] are absorbed into the 
two-particle backward and forward  scatterings by introducing\cite{Giamarchi88}
\be
\tilde g_1&=&g_1-\tilde D_{\xi},\\
\tilde g_2&=&g_2-\tilde D_{\eta},
\en
where 
 $\tilde D_{\xi}={D_{\xi} \Lambda/ \pi v_F^2}$
 and $\tilde D_{\eta}={D_{\eta} \Lambda/ \pi v_F^2}$.
Now, in addition to the two-particle scattering
vertices [Figs.~1(a)-(c)], 
there appear inter-replica vertices as shown in Figs.~6(a) and (b).

\subsection{One-loop renormalization}

We obtain the vertex correction diagrams for 
$\tilde g_1$, $\tilde g_2$ and $g_3$ 
merely by replacing $g_1$ and $g_2$ in  Fig.~2
with $\tilde g_1$ and $\tilde g_2$, respectively.
However, 
we must avoid counting the diagram as shown in Fig.~6(c) which apparently renormalizes
$\tilde g_1$, but vanishes in the replica limit, $N\to 0$,
since summation over the replica indices $\gamma=1,2,...,N$ of the inner loop yields $N$. Keeping this point in mind,
we obtain
 the renormalized scattering strength, $\tilde g'_1$, $\tilde g'_2$, and $g'_3$,
 which are analogous to equations (\ref{vc1})-(\ref{vc3})  and given by
\be
\tilde g'_1&=&\tilde g_1
+{\tilde g_1\tilde g_2}\ln{\omega\over E_0}
-[(\tilde g_2-\tilde g_1)\tilde g_1+\tilde D_{\xi}^2]\pi(\omega)\label{vcr1},\\
\tilde g'_2&=&\tilde g_2+{1\over2}({\tilde g_1^2+\tilde g_2^2})\ln{\omega\over E_0}
-{1\over2}({\tilde g_2^2+g_3^2})\pi(\omega),\\
g'_3&=&g_3+{g_3}(\tilde g_1-2\tilde g_2)\pi(\omega). \label{vcr3}
\en

 Renormalization of 
 the inter-replica vertices   for  $\beta\neq\alpha$  in Figs.~6(a) and (b)
 comes from the vertex correction diagrams as shown in Figs.~7(a) and (b),
 respectively.
We must avoid counting the diagram  Fig.~7(b-4) which
 vanishes in the replica limit.
We obtain
\be
\tilde D_{\eta}'&=&\tilde D_{\eta}-{1\over2}\tilde D_{\xi}^2\ln{\omega\over E_0}\label{vceta},\\
\tilde D_{\xi}'&=&\tilde D_{\xi}-\tilde D_{\eta}\tilde D_{\xi}\ln{\omega\over E_0}
+[(2\tilde g_1-\tilde g_2)\tilde D_{\xi}+2\tilde D_{\xi}^2]\pi(\omega)\label{vcxi}.
\en
When we differentiate (\ref{vcr1})-(\ref{vcxi}) with respect to the scaling parameter,
$l=\ln(E_0/\omega)$,  the length-scale, $\Lambda$, must also be scaled  in accordance
with the change of the energy scale, $\omega$, as
\be
{d\Lambda\over \Lambda}+{d\omega\over \omega}=0.
\en
Thus we obtain the RG equations
\begin{eqnarray}
{\dd  \tilde g_{1}\over  \dd l} &=&-\tilde D_{\xi}-
\tilde  g_{1}\tilde g_{2} +[ (\tilde g_{2}-\tilde  g_{1} ) \tilde  g_{1}+\tilde D_{\xi}^2 ]\lambda_l,\label{eqn:rrgeqg1}\\
{\dd  \tilde g_{2} \over \dd l} &=&-\tilde D_{\eta}-( \tilde g_{1}^2+ \tilde g_{2}^2)/2 +( \tilde g_{2}^2+ g_{3}^2)\lambda_l/2  ,\\
{\dd   g_{3} \over \dd l} &=&-  g_{3} ( \tilde g_{1}-2 \tilde g_{2} ) \lambda_l  \label{eqn:rrgeqg3},\\
{\dd \tilde D_{\eta}\over \dd l}&=&\tilde D_{\eta}+\tilde D_{\xi}^2/2,\\
{\dd \tilde D_{\xi}\over  \dd l} &=&\tilde D_{\xi}+\tilde D_{\xi}\tilde D_{\eta}
-[(2\tilde  g_{1}-\tilde g_{2})\tilde D_{\xi} +2\tilde D_{\xi}^2]
\lambda_l .\label{eqn:rrgeqdx1}
\end{eqnarray}

\subsection{Renormalization-group flow and QPT}
\subsubsection{In the absence of the random backward scattering}
First, we consider the case where 
the random forward scattering is present ($\tilde D_{\eta;0}\neq 0$), but
the random backward scattering is absent ($\tilde D_{\xi;0}=0$), where
$\tilde D_{\eta;0}$ and $\tilde D_{\xi;0}$ are initial strengths of the random 
forward and backward scatterings, respectively.
In this case, the RG flows indicate that the QPT occurs  from
a metallic fixed point, $(\tilde D_{\eta}^\ast=\infty, g_3^\ast=0)$, 
to the Mott insulator fixed point, $(\tilde D_{\eta}^\ast=\infty, g_3^\ast=\infty)$,
as $\tilde U$ increases.
Typical flows are shown in Figs.~8(a-1) and (a-2) for
$\tilde U=0.1$ and 0.4, respectively, in the case of  $d=1.1$ and
$\tilde D_{\eta;0}=0.02$, where the critical value of $\tilde U$
is $\tilde U_{\rm c}\sim 0.330$.
This behavior in $d=1+\epsilon$ qualitatively reproduces  the case of 
$d=1$,\cite{Fujimoto96} where sufficiently strong
 random forward scattering  destroys the Mott gap.

In Fig.~9(a) is shown a low-energy asymptotic phase diagram, where we also show
 energy scales of the Mott gap, $\omega_{\rm gap}/E_0$, introduced 
 in the previous section.
We see that 
both the random forward scattering and
the raising dimensionality  tend to destroy the Mott gap
and consequently widen the metallic region as compared 
with the pure case
[the phase boundary in the pure case is shown by  the gray solid line].

\subsubsection{Effects of the random backward scattering}
Next, we consider the case where both
the random forward and backward scatterings are present:
$\tilde D_{\eta;0}\neq 0$ and $\tilde D_{\xi;0}\neq 0$.
The random backward scattering makes it possible for  the Anderson localization to occur.
In this case, there occurs a transition from
the Anderson insulator fixed point, $(\tilde D_{\xi}^\ast=\infty,\tilde D_{\eta}^\ast=\infty, g_3^\ast=0)$,
to the Mott insulator fixed point, $(\tilde D_{\xi}^\ast=\infty,\tilde D_{\eta}^\ast=\infty,  g_3^\ast=\infty)$,
as $\tilde U$ increases.
Typical flows are shown in Figs.~8(b-1) and (b-3)
 for $\tilde U=0.1$ and 0.4, respectively.
We here introduce the scale $l_{\rm loc}$ at which $\tilde D_{\xi}$ 
reaches unity [see Fig.~8(b-1)].
Then, $\omega_{\rm loc}=E_{0}e^{-l_{\rm loc}}$
gives a qualitative energy scale
around which a crossover to the Anderson insulator occurs.
In the flows of the type of Fig.~8(b-3),  $g_3$ always dominates $\tilde D_{\xi}$
and reaches unity at the scale $l=l_{\rm gap}<l_{\rm loc}$, 
indicating that  the Mott gap formation overwhelms the Anderson localization.

We  also find 
 flows toward $(\tilde D_{\xi}^\ast=0,\tilde D_{\eta}^\ast=\infty, g_3^\ast=0)$,
as shown in Fig.~8(b-2).
This type of flows is found only for $0<d<1.575$ in the  narrow region of $\tilde U$
in between the regions corresponding to Figs.~8(b-1) and 8(b-3).  
In these cases, however,  $\tilde D_{\xi}$ exceeds unity at some scaling
parameter  $l_{\rm loc}$   during the renormalization,
which indicates that the perturbative treatment breaks down and the localization occurs 
around the energy scale specified by $l=l_{\rm loc}$.
Thus we interpret  that the ground state corresponding 
to this fixed point is   the Anderson insulator.
It is beyond the RG-based scheme to settle this ambiguity and we do not go 
into the details on this issue here.

In Fig.~9(b) is shown a low-energy asymptotic phase diagram, where we also
 show $\omega_{\rm loc}/E_0$ and $\omega_{\rm gap}/E_0$.
As compared with Fig.~9(a), the phase boundary remains nearly unchanged,
but the metallic phase in Fig.~9(a) is replaced with the Anderson insulator phase due to the random backward scattering.

The present
results  indicate  that the QPT from the Anderson to the Mott insulators occurs in both $d=1$ and $d>1$.
Recently, Ohtsuka and Hatsugai\cite{OhtsukaHatsugai99} studied  
the half-filled Hubbard model containing  site randomness by
 using the Monte Carlo method. They found  that the 
 QPTs from an incompressible (Mott) to a compressible (Anderson)
 insulators occur  in all the cases of $d=1,2,3$.
This numerical result is consistent with the RG-based views given here.

\section{CONCLUDING REMARKS}

In this paper, based on the one-loop RG flows, 
we have studied  QPTs in the half-filled Hubbard model
with a  circular Fermi surface
in $d=1+\epsilon$ continuous dimensions.
Peculiarity of  the $d=1+\epsilon$ dimensions
was  incorporated only through  the  mathematical structure of the elementary PP and PH loops:
infrared logarithmic singularity of the PH loop is smeared.
We have studied the following three cases:

\noi
(1) {\it In the absence of randomness}: 
The QPT from the metallic phase to the Mott insulator phase
occurs  at a finite $U$ for  $\epsilon> 0$.

\noi
(2) {\it In the case where the random forward scattering is present, but
the random backward scattering is absent}:
 Both random forward scattering  and
 raising dimensionality tend to destroy the Mott gap.
 Consequently, $\tilde U_{\rm c}$ becomes finite  for  $\epsilon\geq 0$
 and the metallic region becomes wider as compared with the pure case (1).

\noi 
(3) {\it
In the case where both the random forward and backward scatterings are
present}:
The phase boundary  remains nearly unchanged as compared with the case (2),
but the metallic phase in the case (2) 
is replaced with the Anderson insulator phase due to the random backward scattering.

In the present study, 
the ground state properties  were conjectured based solely  on the one-loop RG flows.
At present, numerical studies are in progress to complement the views given
 here.\cite{YonemitsuKishine}

\acknowledgments
The author acknowledges Professor K. Ueda for directing his attention to Ref.~[16].
He also thanks Professors K. Yonemitsu and Y. Hatsugai for discussions and useful information.
He was   supported by a Grant-in-Aid for Encouragement of Young
Scientists   from the Ministry of Education, Science, Sports and Culture,
Japan. 

\appendix
\section{Derivation of (10)}
Our purpose here is to show that the $\tilde \omega$-dependence of
$\Re\Pi_{2k_F}(\omega)$ is uniquely determined as
(\ref{eqn:phloop}).
We start with  equation (\ref{eqn:ret}).
The imaginary counterpart is given by
\be
\Im\Pi_{2k_F}(\omega)=-\pi{S_{d-1}\over(2\pi)^d}\int k^{d-1}dk
\int_{0}^\pi d\theta (\sin\theta)^{d-2}
\Theta(-\xi_\k)\Theta(\xi_{\k+\SLQ})\delta(\omega-\xi_{\k+\SLQ}+\xi_\k),
\label{eqn:imag}
\en
for $0<\omega<2v_Fk_F$, 
$\Im\Pi_{2k_F}(\omega)=0$ for $2v_Fk_F<\omega$
and satisfies
$
\Im\Pi_{2k_F}(-\omega)=-\Im\Pi_{2k_F}(\omega).
$
The delta function in the integrand of (\ref{eqn:imag}) is rewritten as
$
\delta(\omega-\xi_{\k+\q}+\xi_\k)=
{1\over v_Fk_F}\left({1\over 2}+{\tilde \omega \over\tilde k}\right)\delta(t-t_0),
$
where we introduced 
$\tilde \omega=\omega/2v_Fk_F$, $\tilde k=k/k_F$, $t=\cos\theta$ and
$t_0=\tilde \omega-(1-\tilde \omega^2)/ \tilde k$.
So we have
\be
\Im\Pi_{2k_F}(\omega)&=&-\pi{S_{d-1}k_F^{d-1}\over (2\pi)^dv_F}
\int_{1-\tilde \omega}^{1} d\tilde k  
 \left({1\over 2}+{\tilde \omega \over\tilde k}\right)
 \tilde k^{d-1}[1-(\tilde \omega-{1-\tilde \omega^2\over\tilde k})^2]^{d-3\over 2}\non\\
&=&
-\pi{S_{d-1}k_F^{d-1}\over 2(2\pi)^dv_F}(1-\tilde \omega^2)^{d-3\over 2}\int_{1}^{1+\tilde \omega} du  
 \left(u^2-\tilde\omega^2\right)(u^2-1)^{d-3\over 2},\label{eqn:exact}
\en
where $u=\tilde k+\tilde\omega$.
For small $\tilde \omega$,
since 
the  region of integration is limited to the vicinity of $u=1$,
it is reasonable to replace
the integrand, $\left(u^2-\tilde\omega^2\right)(u^2-1)^{d-3\over 2}$, 
with
$2^{d-3\over 2} (u-1)^{d-3\over 2}$ and  
we have
\be
\Im\Pi_{2k_F}(\omega)&\sim&-\pi
{S_{d-1}k_F^{d-1}\over 2(2\pi)^dv_F}2^{d-3\over 2}(1-\tilde \omega^2)^{d-3\over 2}
\int_{1}^{1+\tilde \omega} du  (u-1)^{d-3\over 2}\non\\
&=&-\pi{S_{d-1}k_F^{d-1}\over 2(2\pi)^dv_F}{2^{d-1\over 2}\over d-1}
(1-\tilde \omega^2)^{d-3\over 2}\tilde \omega^{d-1\over 2},
\en
which corresponds to
a special case  of  equation (7.12) in Ref.~[17] for $\mid\!\q\!\mid=2k_F$. 
In $d=1+\epsilon$,
noting $S_{d-1}\sim\epsilon$, we obtain
\be
\Im\Pi_{2k_F}(\omega)\sim- 
{1\over 4  v_F}(1-\tilde \omega^2)^{-1+{\epsilon/2}}
\tilde \omega^{\epsilon/2}.\label{eqn:approx}
\en

The real counterpart   is obtained by Kramers-Kronig transformation,
\be
\Re\Pi_{2k_F}(\omega)&=&{2\over\pi}{\cal P}\int_{0}^{1}
d\tilde\omega'{\tilde \omega' \Im\Pi_{2k_F}(\omega')\over \tilde \omega'^2-\tilde \omega^2},\label{eqn:KKo}
\en
where the symbol ${\cal P}$ denotes 
Cauchy principal value integral [Note that $0<\tilde \omega<1$]. 
Since the prefactor, $(1-\tilde \omega^2)^{-1+{\epsilon/2}}$, on the r.h.s of  (\ref{eqn:approx}) has already appeared
in the exact expression, (\ref{eqn:exact}),
and the remaining part of the integral, (\ref{eqn:exact}), shows no singularity at $\tilde\omega=1$, 
the expression for small $\tilde\omega$,
  (\ref{eqn:approx}),  holds 
 analytical property  of $\Im\Pi_{2k_F}(\omega)$ correctly even for $\tilde\omega{\sim}1$.
Thus it is reasonable to  use the expression  (\ref{eqn:approx}) in (\ref{eqn:KKo}) and we have
\be
\Re\Pi_{2k_F}(\omega)
\sim-
{1\over 2\pi v_F}
{\cal P}\!\!\int_{0}^{1}d\tilde\omega'
{\tilde \omega'^{1+\epsilon/2}\over (\tilde \omega'^2-\tilde \omega^2)(1-\tilde\omega'^2)^{1-\epsilon/2}}\label{eqn:KK},
\en
which is to be evaluated   for small 
$\tilde\omega$ and $\epsilon$.

To evaluate  (\ref{eqn:KK}),
let
\begin{eqnarray}
f(z)
={z\over  (z^2-\tilde \omega^2)(z^2-1)}[z(z^2-1)]^{\epsilon/2},
\end{eqnarray}
and consider the integral, $\oint_C f(z) dz$,
along the contour as depicted in Fig.~10.
$f(z)$ has poles at $z=\pm \tilde\omega$ and branch points at $z=0,\infty$ and $z=\pm 1$.
We choose branch cuts in the region $\{\Re z<-1\}\cup \{0<\Re z\}$.
In the limit as the large circle recedes to infinity, it gives no contribution.
The residue   at the pole $z=-\tilde\omega$ gives
\be
\oint_C f(z) dz= -\pi\ii e^{\pi\ii\epsilon/ 2}
{\tilde\omega^{\epsilon/2}\over (1-\tilde\omega^2)^{1-\epsilon/2}} 
\sim-\pi\ii \tilde\omega^{\epsilon/2}.\label{r1}
\en
The remainder of the contour is deformed into an integral enclosing the cuts and encircling 
the pole
$z=\tilde\omega$ and the branch points at $z=0$ and $z=\pm 1$.
 The points $z=0,\pm1$  
give no contribution.
The integrals encircling the pole $z=\tilde\omega$ and  the
remainder along the real axis give 
\be
&&\oint_C f(z) dz
=
{\pi\ii\over 2}(e^{{\pi\over 2}\ii\epsilon}+e^{{3\pi\over 2}\pi\ii\epsilon})
{ \tilde\omega^{\epsilon/2}\over(1-\tilde\omega^2)^{1-\epsilon/2}}
+
2\pi v_F (e^{{\pi\over 2}\ii\epsilon}-e^{{3\pi\over 2}\ii\epsilon})\Re\Pi_{2k_F}(\omega) \non\\
&+&
(e^{{4\pi\over 2}\ii\epsilon}-1)
\ds\int_{1}^{\infty}
{x[x(x^2-1)]^{\epsilon/2}\over(x^2-\tilde\omega^2)(x^2-1)}dx
+(e^{{3\pi\over 2}\ii\epsilon}-e^{{\pi\over 2}\ii\epsilon})\ds\int_{-\infty}^{-1}
{x[-x(x^2-1)]^{\epsilon/2}\over(x^2-\tilde\omega^2)(x^2-1)}dx.\non
\en
Here the integrals in the second line are evaluated
for small $\tilde\omega$ and $\epsilon$ as  
\be
&&\int_{1}^{\infty}
{x[x(x^2-1)]^{\epsilon/2}\over(x^2-\tilde\omega^2)(x^2-1)}dx
=-\int_{-\infty}^{-1}
{x[-x(x^2-1)]^{\epsilon/2}\over(x^2-\tilde\omega^2)(x^2-1)}dx\non\\
&\sim& \int_{1}^{\infty}
[x(x^2-1)]^{-1+\epsilon/2}dx
={\Gamma(1-3\epsilon/4)
\Gamma(\epsilon/2)\over2\Gamma(1-\epsilon/2)}\sim 1/\epsilon,\non
\en
where the integral in the second line
converges for $0<\epsilon<4/3$. 
We thus obtain, for small $\tilde \omega$ and $\epsilon$,
\be
\oint_C f(z) dz=
\pi\ii \tilde\omega^{\epsilon/2}
-
2\pi^2 v_F \ii\epsilon\Re\Pi_{2k_F}(\omega)
+2\pi\ii-\pi\ii.\label{pole}
\en
Therefore we   obtain $\Re\Pi_{2k_F}(\omega)$ of the form
which is correct up to the leading order of $\tilde\omega$,
\be
\Re\Pi_{2k_F}(\omega)\sim{1\over 2\pi v_F}
\left[{ \tilde\omega^{\epsilon/2}\over \epsilon/2}+C_\epsilon\right],\label{result}
\en
which is (\ref{eqn:phloop}).
Although above manipulation   gives $C_\epsilon=1/\epsilon$, 
  it seems feasible to choose $C_\epsilon=-2/\epsilon$ as suggested in Ref.~[16]
to  reproduce correctly 
the limit form of $\Re\Pi_{2k_F}(\omega)={1\over 2\pi v_F}
\log\tilde\omega$ at $\epsilon= 0$.
This discrepancy may arise, because to evaluate (\ref{eqn:KKo}) we used  
the expression  (\ref{eqn:approx}) which   holds 
 analytical property  of $\Im\Pi_{2k_F}(\omega)$ correctly   
but misses contribution from $\tilde\omega{\sim}1$.
We do not   go into the details   here, since
 an explicit form of $C_\epsilon$ does not enter
the RG equations (\ref{eqn:rgeqg1})-(\ref{eqn:rgeqg3})
and (\ref{eqn:rrgeqg1})-(\ref{eqn:rrgeqdx1}).

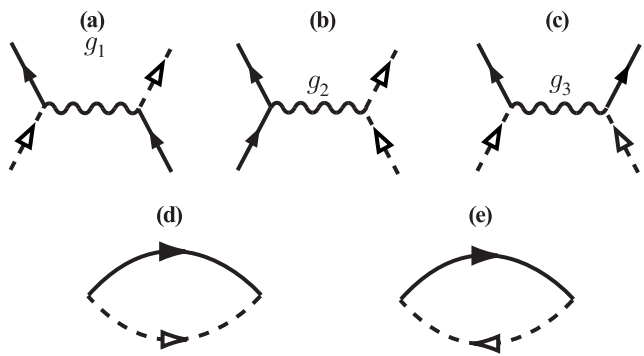
\begin{figure}
\caption{
Two-particle scattering
vertices for the (a) backward, (b) forward, and (c) $2k_F$-umklapp
scattering processes,
and the elementary (d) particle-particle (PP) and (e) particle-hole (PH) loops.}
\end{figure}

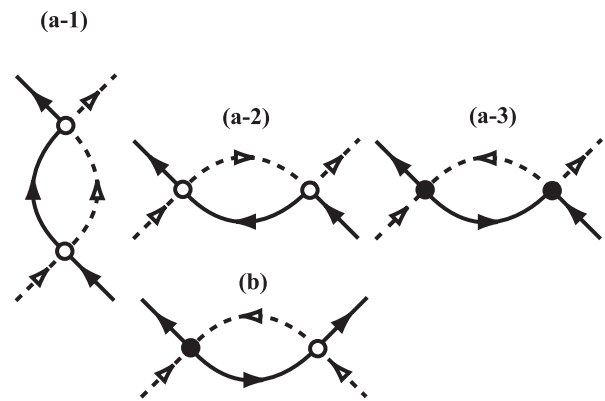
\begin{figure}
\caption{Vertex correction diagrams for
the (a) normal [$g_1$, $g_2$] and (b)
umklapp [$g_3$] processes.
} White and black circles represent the normal and umklapp scatterings, respectively.
\end{figure}

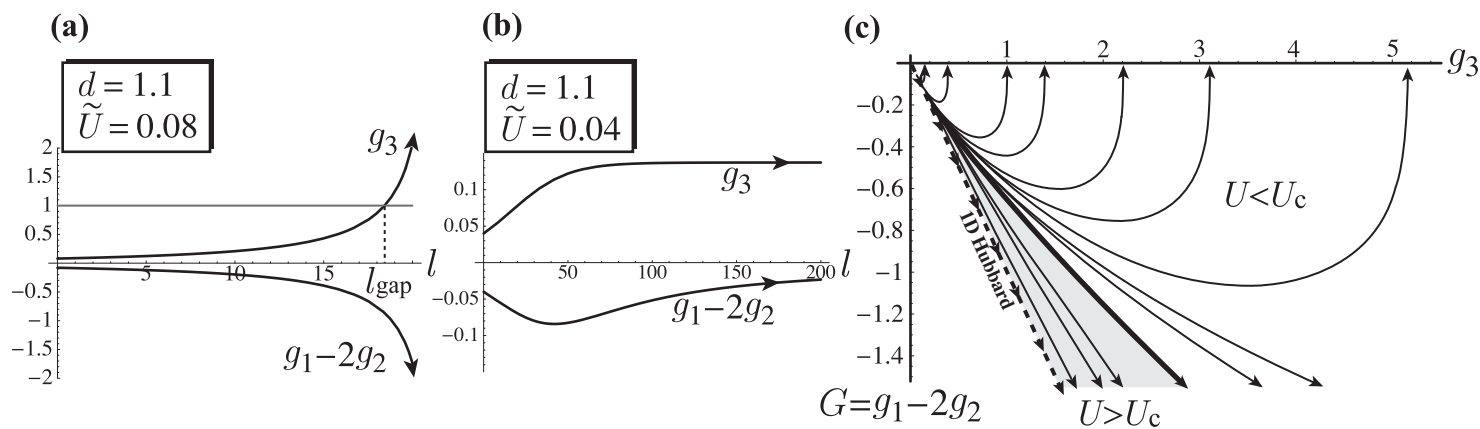
\begin{figure}
\caption{
 RG flows in the case of $d=1.1$ with 
(a) $\tilde U=0.08$ and (b) $0.04$.
(c) The RG trajectories in terms of  $g_3$ and $G=g_1-2g_2$ in $d=1.1$. 
A critical value is  $\tilde U_{\rm c}=0.0588$.
In the $d=1$ half-filled Hubbard model, 
$g_3$ and $G$ flow along the  broken line denoted by \lq\lq 1D Hubbard\rq\rq.}
\end{figure}

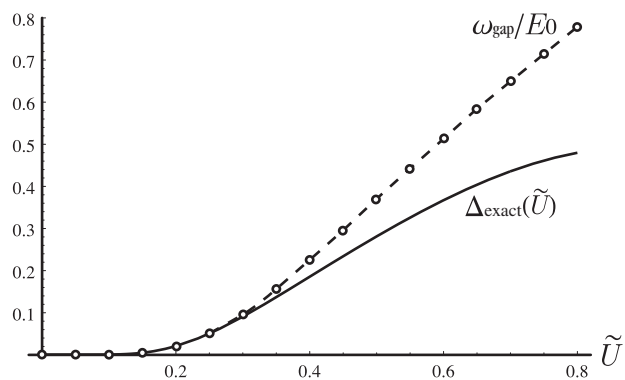
\begin{figure}
\caption{
 $\tilde U$-dependence of 
$
\Delta_{\rm exact}(\tilde U)
$
and  $\omega_{\rm gap}=E_{0}e^{-l_{\rm gap}}$ in $d=1$.
}
\end{figure}

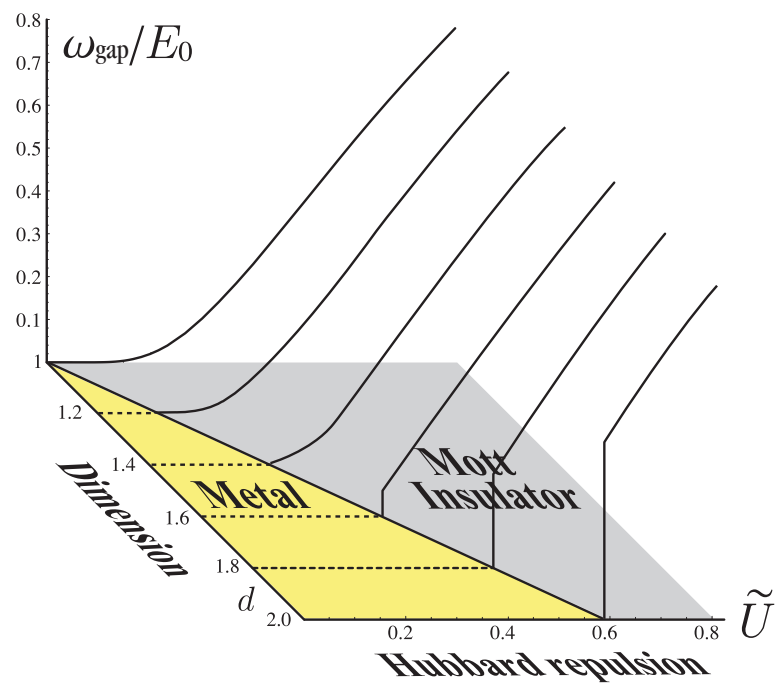
\begin{figure}
\caption{A low-energy asymptotic phase diagram of the $d=1+\epsilon$ dimensional
 Hubbard model at half filling.
We also show  energy scales of the Mott gap,
$\omega_{\rm gap}/E_0$, as a function of $d$ and $\tilde U$.
}
\end{figure}

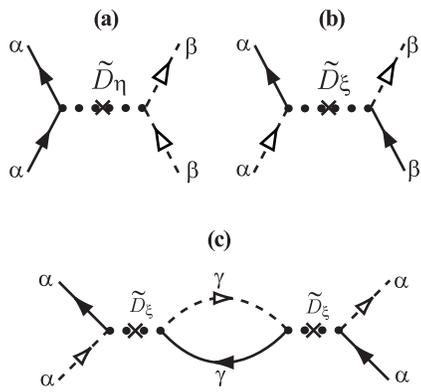
\begin{figure}
\caption{
Inter-replica vertices originating from
the random (a) forward and (b) backward scatterings.
(c) The diagram  which
 is proportional to the number of replicas, $N$, and   
 vanishes in the replica limit, $N\to 0$.}
\end{figure}

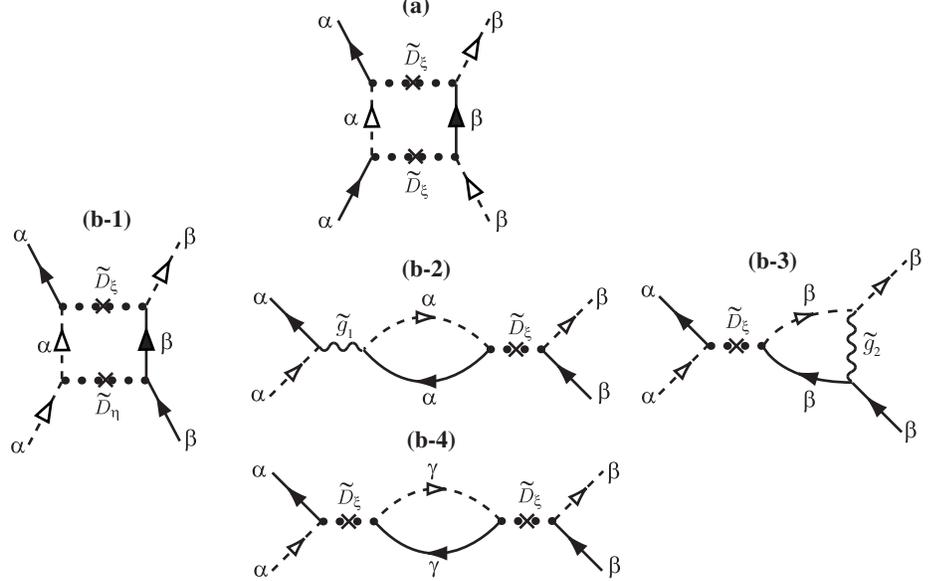
\begin{figure}
\caption{
The vertex correction diagrams for  the inter-replica
(a) forward [$\tilde D_{\eta}$] and (b) backward [$\tilde D_{\xi}$]
processes.
We must avoid counting the diagram (b-4) which
 vanishes in the replica limit.
}
\end{figure}

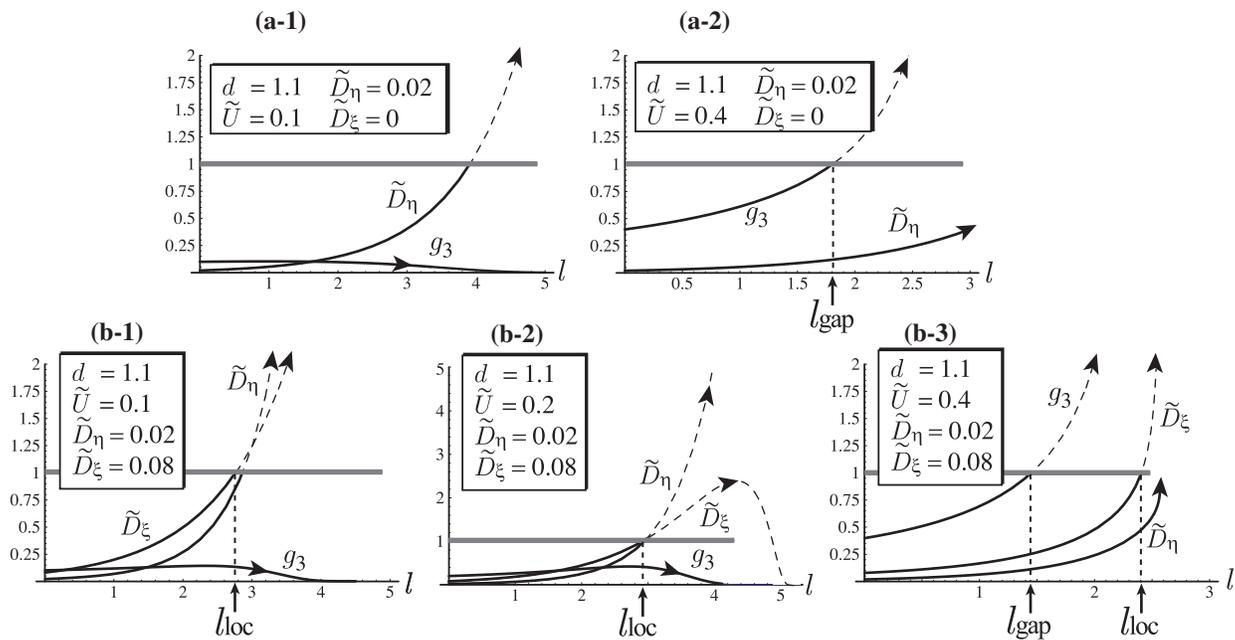
\begin{figure}
\caption{
(a) RG flows of $\tilde D_{\eta}$ and $g_3$
 for
(a-1) $\tilde U=0.1$ and (a-2) $\tilde U=0.4$ in the case of  $d=1.1$,  
$\tilde D_{\eta;0}=0.02$  and $\tilde D_{\xi;0}=0$.
(b) RG flows of $\tilde D_{\eta}$, $\tilde D_{\xi}$ and $g_3$
 for
(b-1) $\tilde U=0.1$, (b-2) $\tilde U=0.2$ and (b-3) $\tilde U=0.4$ in the case of  $d=1.1$,  
$\tilde D_{\eta;0}=0.02$ and $\tilde D_{\xi;0}=0.08$.
}
\end{figure}

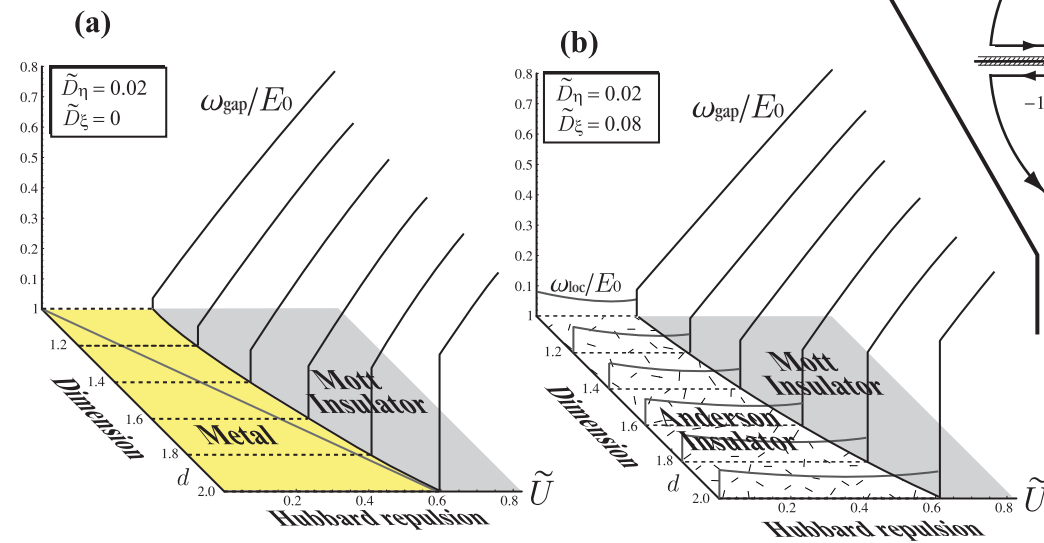
\begin{figure}
\caption{
Low-energy asymptotic phase diagrams 
of the $d=1+\epsilon$ dimensional random Hubbard model at half filling in  
 the cases where
(a) the random forward scattering is present, 
but the random backward scattering is absent,
and
(b) both the random forward and backward scatterings are present.
In (a), the phase boundary in the pure case [see Fig.~5] is shown by the
gray solid line.
We also show $\omega_{\rm loc}/E_0$ and $\omega_{\rm gap}/E_0$,  as  a function of $d$ and $\tilde U$.
}
\end{figure}

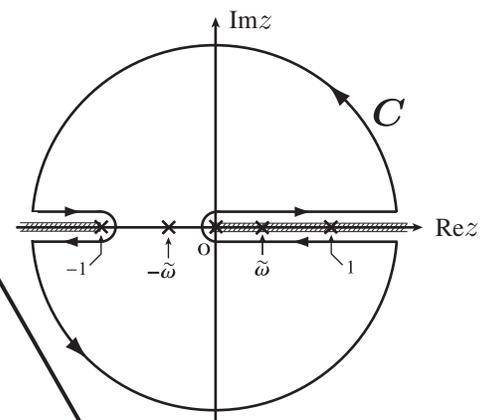
\begin{figure}
\caption{
Contour  to evaluate  the integral (A6).
}
\end{figure} 

\end{document}